\documentclass[conference]{IEEEtran}
\IEEEoverridecommandlockouts
\IEEEpubid{\makebox[\columnwidth]{© 2025 NVIDIA. All rights reserved. \hfill}
\hspace{\columnsep}\makebox[\columnwidth]{ }}
\usepackage[nolist]{acronym} 
\usepackage{cite}
\usepackage{comment}
\usepackage{amsmath,amssymb,amsfonts}
\usepackage{algorithmic}
\usepackage{graphicx}
\usepackage{comment}
\usepackage{textcomp}
\usepackage{xcolor}
\usepackage{floatrow} 
\usepackage{wrapfig}
\usepackage{color}
\usepackage{verbatim}
\usepackage{adjustbox}
\usepackage{hyperref}
\usepackage{tabularx}

\usepackage{url}

\usepackage{subfigure}
\usepackage{booktabs}
\usepackage{textgreek}
\usepackage{enumitem}
\usepackage[perpage]{footmisc}

\usepackage{listings}

\def\BibTeX{{\rm B\kern-.05em{\sc i\kern-.025em b}\kern-.08em
    T\kern-.1667em\lower.7ex\hbox{E}\kern-.125emX}}

\newcommand{\openshmem}[1][]{%
 {Open\-SHMEM\ifthenelse{\equal{#1}{}}{}{~#1}}}

\begin{acronym}
\acro{PVM}{Parallel Virutal Machine}
\acro{PICL}{Portable Instrumented Communication Library}
\acro{SSID}{Synchronization Sequence Identification}
\acro{phit}{physical units}
\acro{BTE}{Byte Transfer Engine}
\acro{FMA}{Fast Memory Access}
\acro{MPI}{\emph{Message Passing Interface}}
\acro{UPC}{\emph{Unified Parallel C}}
\acro{PGAS}{Partitioned Global Address Space}
\acro{SIMD}{\emph{Single Instruction Multiple Data}}
\acro{MIMD}{\emph{Multiple Instruction Multiple Data}}
\acro{HPC}{\emph{High Performance Computing}}
\acro{API}{\emph{Application Programming Interface}}
\acro{RDMA}{\emph{Remote Direct Memory Access}}
\acro{AMO}{\emph{Atomic Memory Operations}}
\acro{CRC}{\emph{Cyclic Redundancy Check}}
\acro{MDH}{\emph{Memory Descriptor Handle}}
\acro{PE}{\emph{Programming Element}}
\acro{ORNL}{Oak Ridge National Laboratory}
\acro{UNM}{University of New Mexico}
\acro{MPI}{Message Passing Interface}
\acro{PGAS}{Partioned Global Address Space}
\acro{HPC}{High Performance Computing}
\acro{UCCS}{Universal Common Communication Substrate}
\acro{PROMPI}{Protocol Reconfiguration and Optimization system for \ac{MPI}}
\acro{LANL}{Los Alamos National Laboratory}
\acro{DOE}{Department of Energy}
\acro{DoD}{Department of Defense}
\acro{OLCF}{\emph{Oak Ridge Leadership Computing Facility}}
\acro{QMC}{\emph{Quantum Monte Carlo}}
\acro{CPU}{\emph{Central Processing Unit}}
\acro{SpT}{QMC Samples per Thread}
\acro{AMR}{\emph{Adaptive Mesh Refinement}}
\acro{WL-LSMS}{\emph{Wang Landau - Locally Self-consistent Multiple Scattering}}
\acro{WL}{\emph{Wang Landau}}
\acro{LSMS}{\emph{Locally Self-consistent Multiple Scattering}}
\acro{rLIZ}{\emph{Local Interaction Zone Radius }}
\acro{LAMMPS}{\emph{Large-scale Atomic/Molecular Massively Parallel Simulator}}
\acro{NRDF}{\emph{Non-equilibrium Radiation Diffusion}}
\acro{CESM}{\emph{Community Earth System Model}}
\acro{Vampir}{\emph{Vampir Tool Suite}}
\acro{SAMRAI}{\emph{Structured Adaptive Mesh Refinement Application Infrastructure}}
\acro{CHARMM}{Chemistry at HARvard Macromolecular Mechanics}
\acro{PPPM}{particle-particle particle-mesh}
\acro{IOFSL}{I/O Forwarding Scalability Layer}
\acro{PE}{Processing Element}
\acro{APP}{Application Time}
\acro{COMP}{Computation Time}
\acro{COL}{Collective}
\acro{GPU}{Graphical Processing Unit}
\acro{SIMD}{\emph{Single Instruction Multiple Data}}
\acro{MIMD}{\emph{Multiple Instruction Multiple Data}}
\acro{API}{\emph{Application Programming Interface}}
\acro{RDMA}{\emph{Remote Direct Memory Access}}
\acro{UPC}{\emph{Unified Parallel C}}
\acro{SPMD}{\emph{Single Program Multiple Data}}
\acro{MTTI}{Mean Time Between Interrupts}
\acro{CAF}{Co-Array Fortran}
\acro{SHMEM}{Symmetric Hierarchical Memory}
\acro{FLOPS}{FLoating-point Operations Per Second} 
\acro{MPMD}{Multiple Program Multiple Data} 
\acro{SM}{Shared-Memory} 
\acro{OS}{Operating System} 
\acro{UH}{University of Houston} 
\acro{SNL}{Sandia National Laboratory} 
\acro{ANL}{Argonne National Laboratory} 
\acro{NOP}{No Operation} 
\acro{NUMA}{Non Uniform Memory Access}
\acro{BTL}{Byte Transfer Layer}
\acro{PML}{Point-to-point Management Layer}
\acro{BML}{BTL Management Layer}
\acro{MCA}{Modular Component Architecture}
\acro{DAG}{\emph{Directed Acyclic Graph}}
\acro{BCOL}{\emph{Basic Collective}}
\acro{ML}{\emph{Messaging Layer}}
\acro{P2P}{\emph{Point-to-Point}}
\acro{SM}{\emph{Shared-Memory}}
\acro{MB}{\emph{Megabyte}}
\acro{ORTE}{Open Run-time Environment}
\acro{OPAL}{Open Portable Access Layer}
\acro{IBM}{International Business Machines Corportation}
\acro{HP}{Hewlett-Packard}
\acro{SGI}{Silicon Graphics, Inc.}
\acro{DMAPP}{Distributed Memory Applications}
\acro{VM}{Virtual Machine}
\acro{TCO}{Total Cost of Ownership}
\acro{EC2}{Elastic Compute Cloud}
\acro{NERSC}{National Energy Research Scientific Computing Center}
\acro{ALCF}{Argonne Leadership Computing Facility}
\acro{MR}{\emph{MapReduce}}
\acro{NIST}{National Institute of Standards and Technology}
\acro{10GbE}{10-gigabit Ethernet}
\acro{CUDA}{\emph{Compute Unified Device Architecture}}
\acro{SMs}{Streaming Multiprocessors}
\acro{FFT}{Fast Fourier transform}
\acro{UCX}{Unified Communication X}
\acro{UCS}{UC-Services}
\acro{UCT}{UC-Transports}
\acro{UCP}{UC-Protocols}
\acro{TL}{Transport Layer}
\acro{RMA}{Remote Memory Access}
\acro{PMI}{Process Manager Interface}
\acro{SMSG}{Short message}
\acro{BTE}{Block Transfer Engine}
\acro{SSCA}{Scalable Synthetic Compact Applications}
\acro{RTE}{Run Time Environment}
\acro{SharP}{SHARed data-structure centric Programming abstraction}
\acro{HBM}{High-bandwidth Memory}
\acro{NVRAM}{non-volatile random access memory}
\acro{LLNL}{Lawerence Livermore National Laboratory}
\acro{MD}{\emph{Memory Domains}}
\acro{DT}{\emph{Data Tiers}}
\acro{UMA}{\emph{Unified Memory Allocator}}
\acro{UMI}{\emph{Unified Memory Interface}}
\acro{KV}{\emph{Key-Value}}
\acro{C/R}{\emph{Checkpoint/Restart}}
\acro{ADIOS}{Adaptable IO System}
\acro{MDS}{Metadata Server}
\acro{TEPS}{Traversed Edges Per Second}
\acro{UCX}{Unified Communication X}
\acro{UCS}{UC-Services}
\acro{UCT}{UC-Transports}
\acro{UCP}{UC-Protocols}
\acro{TL}{Transport Layer}
\acro{TLS}{Thread Local Storage}
\acro{SSCA}{Scalable Synthetic Compact Applications}
\acro{SHARP}{Mellanox Scalable Hierarchical Aggregation and Reduction Protocol}
\acro{NCCL}{NVIDIA Collective Communication Library}
\acro{RCCL}{ROCm Communication Collectives Library}
\acro{ANL}{Argonne National Laboratory}
\acro{LANL}{Los Alamos National Laboratory}
\acro{ORNL}{Oak Ridge National Laboratory}
\acro{SNL}{Sandia National Laboratory}

\acro{HCOLL}{Hierarchical Collectives}
\acro{HCA}{Host Channel Adapter}
\acro{RMA}{Remote Memory Access}
\acro{MEMIC}{Memory Mapped to InterConnect}
\acro{BAR}{Base Address Registers}
\acro{UCC}{Unified Collective Communication}
\acro{DL}{Deep Learning}
\acro{EE}{Execution Engine}
\acro{CL}{Collective Layer}
\acro{TL}{Team Layer}
\acro{DPU}{Data Processing Unit}
\end{acronym}

\begin{document}

\pagestyle{plain}  

\title{GPU-Initiated Networking for NCCL}

    
\author{
\IEEEauthorblockN{
    Khaled Hamidouche, 
    John Bachan,
    Pak Markthub, 
    Peter-Jan Gootzen,
    Elena Agostini,\\
    Sylvain Jeaugey,
    Aamir Shafi,
    Georgios Theodorakis, 
    Manjunath Gorentla Venkata
}\\
\textit{
\small
\{khamidouche,~jbachan,~pmarkthub,~pgootzen,~eagostini,~sjeaugey,~ashafi,~gtheodorakis,~manjunath\}@nvidia.com
}\\
NVIDIA Corporation
}

\maketitle

\begin{abstract}%

Modern AI workloads, especially Mixture-of-Experts (MoE) architectures, 
increasingly demand low-latency, fine-grained GPU-to-GPU communication with 
device-side control. Traditional GPU communication follows a host-initiated model, 
where the CPU orchestrates all communication operations---a characteristic of the 
CUDA runtime. Although robust for collective operations, applications requiring 
tight integration of computation and communication can benefit from device-initiated 
communication that eliminates CPU coordination overhead.

NCCL 2.28 introduces the \textbf{Device API} with three operation modes: Load/Store 
Accessible (LSA) for NVLink/PCIe, Multimem for NVLink SHARP, and GPU-Initiated 
Networking (GIN) for network RDMA. This paper presents the \textbf{GIN} architecture, 
design, semantics, and highlights its impact on MoE communication.
GIN builds on a three-layer architecture: 
i)~NCCL Core host-side APIs for device communicator setup and collective memory window registration; 
ii)~Device-side APIs for remote memory operations callable from CUDA kernels; and
iii)~A network plugin architecture with dual semantics (GPUDirect Async Kernel-Initiated and Proxy) for broad 
hardware support.
The GPUDirect Async Kernel-Initiated backend leverages DOCA GPUNetIO for direct GPU-to-NIC communication, while 
the Proxy backend provides equivalent functionality via lock-free GPU-to-CPU queues 
over standard RDMA networks. We demonstrate GIN's practicality through integration 
with DeepEP, an MoE communication library. Comprehensive benchmarking shows that GIN 
provides device-initiated communication within NCCL's unified runtime, combining 
low-latency operations with NCCL's collective algorithms and production infrastructure.

\end{abstract}%

\begin{IEEEkeywords}
NCCL Device API, GPU-Initiated Networking, Load/Store Accessible, Multimem, 
RDMA, One-Sided Communication, Device-Initiated Communication, DeepEP
\end{IEEEkeywords}

\section{Introduction}
\label{sec:introduction}

The rapid development of large language models (LLMs) has introduced new 
performance demands for GPU communication libraries. Modern AI workloads 
require more than traditional collective communication: they require 
low-latency point-to-point operations for inference token 
generation~\cite{tensorrt-llm,vllm}, custom communication patterns for 
Mixture-of-Experts (MoE) architectures~\cite{deepseekai2024deepseekv3}, 
and close integration of computation and communication in compiler-generated 
kernels~\cite{triton,jax}. These workloads benefit from \emph{GPUs directly 
initiating and controlling network communication without CPU involvement}.

NCCL~\cite{nccl} has established itself as the de facto communication 
runtime for GPU-based machine learning, providing optimized collective 
algorithms and robust infrastructure for distributed training and inference. 
Traditional GPU communication follows a host-initiated model, where the CPU 
orchestrates all communication operations. This approach requires explicit 
host-device synchronization---a characteristic of the CUDA runtime model---and 
separate kernel launches for each communication call. While this model has 
proven robust for large-scale collective operations, applications requiring 
tight integration of computation and communication---such as dynamic token 
routing in MoE inference~\cite{deepseekai2024deepseekv3}, compiler-generated 
communication in JAX~\cite{jax} and Triton~\cite{triton} kernels---require 
device-initiated communication that eliminates the CPU coordination overhead 
stemming from CUDA-based host-side synchronization. On the other hand, the NVSHMEM 
library~\cite{nvshmem, nvidia2023nvshmem} has successfully demonstrated the 
viability and performance impact of GPUDirect Async Kernel-Initiated (GDAKI) 
capabilities, providing device primitives that enable communication and 
computation fusion for AI workloads.

\begin{figure}[t]
\centering
\includegraphics[width=0.95\columnwidth]{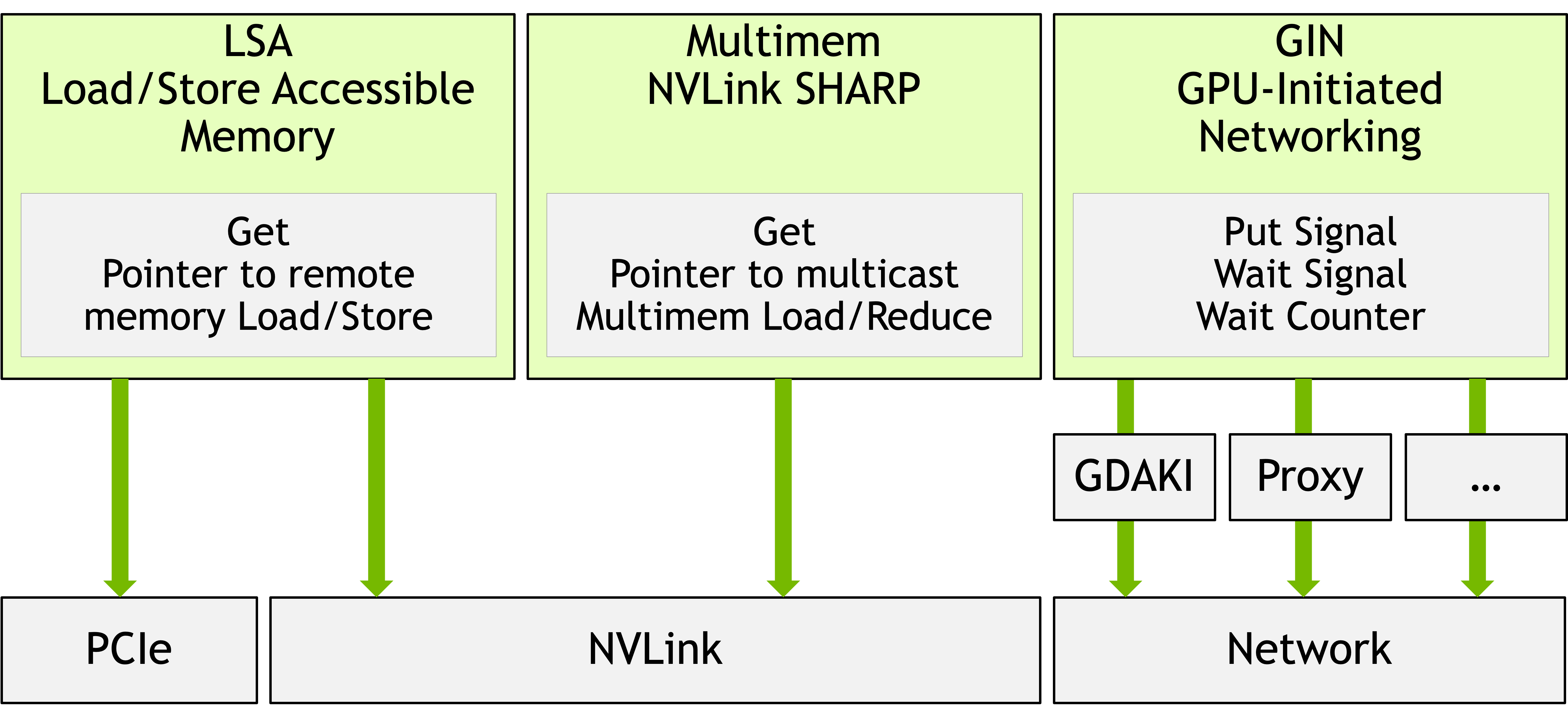}
\caption{NCCL Device API architecture showing three operation modes and their 
underlying interconnect technologies. Load/Store Accessible (LSA) uses PCIe 
and NVLink for intra-node memory operations, Multimem leverages NVLink SHARP 
for hardware multicast, and GPU-Initiated Networking (GIN) provides dual 
backend implementations (GDAKI and Proxy) for network-based communication 
over InfiniBand and RoCE.}
\label{fig:nccl-device-api}
\end{figure}

To satisfy the tight integration of computation and communication---required by modern 
AI workloads---NCCL 2.28 introduces the \textbf{Device API}~\cite{nccl2.28}, 
enabling GPUs to initiate communication operations directly from within kernels. 
The Device API supports three operation modes for device-initiated communication 
(Figure~\ref{fig:nccl-device-api}): \textbf{Load/Store Accessible (LSA)} for 
intra-node communication over NVLink and PCIe using memory load/store 
operations, \textbf{Multimem} for hardware multicast via NVLink SHARP, and 
\textbf{GPU-Initiated Networking (GIN)} for inter-node communication over 
InfiniBand and RoCE networks. This paper focuses on \textbf{GIN}, which 
enables GPUs to initiate network operations directly within a GPU kernel. 
GIN provides device-side APIs for one-sided operations, allowing CUDA kernels 
to perform remote memory operations, point-to-point synchronizations, and poll 
for completion entirely from device code. NCCL's Device API with its GIN 
capabilities provides AI workloads with low-latency primitives, fusion, and 
customization opportunities while enabling these applications to leverage 
NCCL's existing infrastructure, such as hierarchical communicators, elasticity, 
and fault-tolerance mechanisms for large-scale production deployments.

\begin{figure}[t]
\centering
\includegraphics[width=0.95\columnwidth]{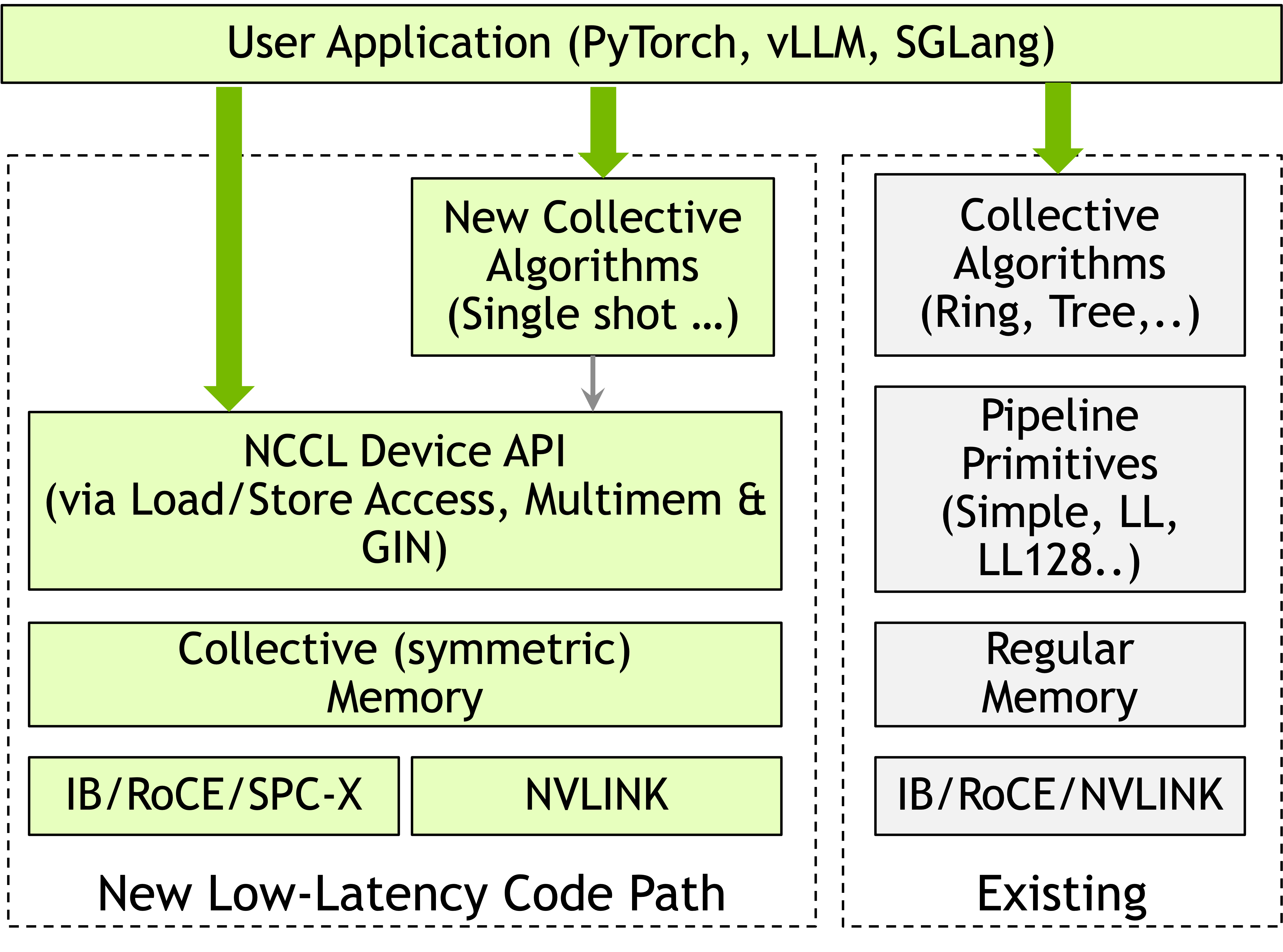}
\caption{High-level architecture comparison: NCCL Device API (left) with 
three operation modes (Load/Store Accessible for NVLink/PCIe, Multimem for 
NVLink SHARP, GIN for Network RDMA) enables single-shot collective algorithms 
with collective symmetric memory, while traditional NCCL (right) uses 
host-initiated algorithms with pipeline primitives over regular memory.}
\label{fig:gin-high-level}
\end{figure}

Figure~\ref{fig:gin-high-level} compares the NCCL Device API architecture 
with traditional host-initiated NCCL. Applications 
(PyTorch~\cite{paszke2019pytorch}, TRT-LLM~\cite{tensorrt-llm}, 
vLLM~\cite{vllm}, SGLang~\cite{sglang}) can either use NCCL-provided 
single-shot collective algorithms implemented with the Device API or directly 
invoke Device API primitives to implement custom communication patterns in GPU 
kernels. The Device API operates over collective symmetric memory, contrasting 
with traditional NCCL's host-initiated collectives that use pipeline primitives 
(Simple, LL, LL128) over regular memory. 

GIN achieves this through a three-layer 
architecture: 
i)~a host-side API extending NCCL Core for communicator 
initialization, GIN resource management, and collective memory window 
registration; 
ii)~a device-side API that exposes \texttt{put}/\texttt{signal} primitives for 
remote memory operations directly callable from CUDA kernels with flexible 
completion semantics; and
iii)~a pluggable network backend architecture supporting both 
direct GPU-to-NIC communication via DOCA GPUNetIO (GDAKI backend) and 
CPU-assisted operation via lock-free queues (Proxy backend).
By providing both hardware-direct and CPU-assisted 
plugin interfaces, GIN offers functionality across diverse deployment scenarios 
while maintaining compatibility with NCCL's existing ecosystem. Unlike typical 
layered architectures, GIN introduces minimal overhead through compile-time 
optimizations and direct hardware access.

\subsection{Key GIN Design Elements}
\label{sec:introduction:design}

The design and implementation of GIN incorporate several key technical elements: 

\begin{enumerate}[label=\roman*)]

    \item\textbf{Device API Design.} GIN provides a device-side API that
    supports flexible cooperation models (thread-level and warp-level 
    collective operations) with optimized primitives for small inline values, 
    window-based memory addressing with byte offsets, and flexible local 
    (flush and counter) and remote completion (signal) mechanisms. This allows 
    applications to express custom communication patterns and fuse 
    communication with computation within kernels.
    
    \item\textbf{Dual Plugin Architecture.} NCCL GIN implements two plugin 
    architectures: the GDAKI interface that leverages the DOCA GPUNetIO 
    backend to enable GPU threads to directly program InfiniBand/RoCE NICs 
    through device verbs, while the Proxy interface uses lock-free GPU-to-CPU 
    queues with 64-byte descriptors to enable GIN 
    functionality on any RDMA-capable NIC.
    
    \item\textbf{Synchronization Mechanisms.} GIN provides local and remote 
    completion notification primitives: signals (for remote 
    notifications) and counters (local completion tracking). These 
    integrate with CUDA's memory model through automatic fence insertion based
    on thread scope annotations.
    
    \item\textbf{Memory Management.} GIN uses collective window registration
    (\texttt{ncclCommWindowRegister}), where each rank contributes a local 
    buffer and receives handles containing remote keys for all peers, enabling 
    zero-copy one-sided operations.

\end{enumerate}

\subsection{Contributions}
\label{sec:introduction:contributions}

In summary, the main contributions of this work are:
\begin{enumerate}[label=\roman*)]
    
    \item Design and implementation of GIN in NCCL, including unified 
        host and device APIs, modular synchronization primitives for 
        asynchronous completion (signals and counters), and two interchangeable 
        backend architectures: GDAKI for direct GPU-to-NIC communication using 
        DOCA GPUNetIO, and Proxy for CPU-assisted operation over standard RDMA.

    \item Integration with DeepEP, a specialized MoE communication library, 
        demonstrating GIN's practical applicability and compatibility with 
        existing NVSHMEM-based device-initiated communication.
    
    \item Comprehensive performance evaluation of GIN using microbenchmarks and 
        application-level experiments with DeepEP kernels, with comparative 
        analysis to establish performance characteristics.

\end{enumerate}

The remainder of this paper is organized as follows.
Section~\ref{sec:background} provides the background on GPU communication and 
motivates the need for GIN.
Section~\ref{sec:nccl-gin} presents the architecture, design principles, 
device-side API semantics, and implementation of GIN—including the three-layer 
design and both GDAKI and Proxy backends, along with synchronization 
primitives and memory management.
Section~\ref{sec:integration} describes the integration of GIN with DeepEP, 
demonstrating practical use for dynamic MoE workloads.
Section~\ref{sec:results} evaluates GIN's performance through microbenchmarks 
and application-level benchmarks in DeepEP across multi-node GPU clusters.
Section~\ref{sec:related} reviews related work in GPU-direct communication and 
one-sided programming models.
Section~\ref{sec:conclusion} concludes with lessons learned and future 
directions for GIN.

\section{Background}
\label{sec:background}

Traditional communication libraries like OpenSHMEM provide one-sided primitives
(\texttt{put}, \texttt{get}, atomics) over symmetric memory regions, enabling
asynchronous data movement without sender-receiver
coordination~\cite{openshmem2012spec}. However, these specifications assume
CPU-centric execution where all communication primitives are invoked from host
code. GPU integration into HPC systems exposed inefficiencies in this model:
fine-grained GPU-to-GPU communication incurred kernel launch overhead, PCIe
transfers to stage data through host memory, and CPU scheduling
latency~\cite{potluri2017gpu}. These bottlenecks motivated GPU-aware extensions
that enable device-initiated communication directly from CUDA
kernels~\cite{venkata2015exploring,potluri2017gpu}.

\subsection{GPUDirect Technologies}

\textbf{GPUDirect RDMA}~(2013)~\cite{nvidia2013gpudirect,potluri2013efficient} 
enables RDMA-capable network interfaces to
directly access GPU memory via PCIe Base Address Register (BAR) mappings,
eliminating CPU and host memory from the data path for inter-node
transfers. The NIC's DMA engine
performs PCIe peer-to-peer transactions to GPU BARs, accessing memory regions
registered through the \texttt{nvidia\_p2p} kernel module. However, GPUDirect
RDMA provides \emph{consistency guarantees only at kernel boundaries}. GPU
memory model semantics (relaxed ordering, write-back caching) prevent safe
concurrent access to RDMA-registered memory from executing kernels, forcing
applications to separate computation and communication~\cite{hamidouche2020gio}.

\textbf{GPUDirect Async}~(2016)~\cite{agostini2018gpudirect} 
introduced partial control-path offload: GPU
threads trigger pre-configured network operations writing to NIC doorbell registers 
that are memory-mapped into the GPU address space.
However, the CPU must pre-construct communication descriptors, limiting
operations to those pre-configured by the host and preventing fully autonomous
device-driven networking.

\subsection{Device-Initiated Communication Primitives}

Fully device-initiated networking requires implementing network programming
interfaces directly in GPU code. Early prototypes like \textbf{GPUrdma}~\cite{daoud2016gpurdma} and
\textbf{GIO}~\cite{hamidouche2020gio} exposed InfiniBand verbs as device-callable functions but faced
GPU-NIC memory consistency challenges.

\textbf{NVSHMEM}~\cite{nvidia2023nvshmem} extended OpenSHMEM semantics to GPU clusters with device-callable
one-sided operations (\texttt{put}, \texttt{get}, atomics) invokable from CUDA
kernels. This enables the device code to interleave
computation and communication without kernel launch overhead, using transport
backends including IBGDA (InfiniBand with GPUDirect Async) for inter-node
transfers and symmetric memory mechanisms for intra-node communication.

\textbf{DOCA GPUNetIO} provides GPU-side RDMA APIs for InfiniBand and RoCE 
networks (IBGDA), exposing device functions that enable GPU kernels to directly program
NICs~\cite{doca-gpunetio}. Specifically, it implements GPUDirect RDMA (direct GPU data movement)
and GPUDirect Async Kernel-Initiated (GPU controlling network communications) technologies.
It forms the foundation for GIN's GDAKI backend, enabling direct GPU-to-NIC communication through
hardware-supported device verbs.

\subsection{Networking Hardware for GPU-Initiated Communication}

GPU-initiated networking requires network interface cards supporting direct
device access through one of several RDMA technologies: InfiniBand, RoCE, or iWARP~\cite{nvidia2022rdmacomparison}.

\textbf{InfiniBand} provides native RDMA support with credit-based flow control,
achieving port-to-port latencies of approximately 130~ns and supporting tens of
thousands of nodes per subnet~\cite{cloudswitch2025ibroce}. InfiniBand adapters
expose memory-mapped queue pairs, completion queues, and doorbell registers
through PCIe BARs, enabling direct GPU access when combined with GPUDirect
RDMA~\cite{potluri2017gpu,nvidia2013gpudirect}.

\textbf{RoCE} implements RDMA over standard Ethernet, offering lower cost and
broader compatibility with existing data center
infrastructure~\cite{cloudswitch2025ibroce,nvidia2022rdmacomparison}. RoCEv2,
the prevalent variant, encapsulates InfiniBand transport over UDP/IP, achieving
port-to-port latencies around 400~ns—higher than native InfiniBand but
sufficient for many workloads~\cite{cloudswitch2025ibroce}. RoCE requires
lossless Ethernet configurations using Priority Flow Control (PFC) and Explicit
Congestion Notification (ECN) to prevent packet loss, which can complicate
deployment in multi-tenant environments~\cite{cloudswitch2025ibroce}. Both
InfiniBand and RoCE share the same user-space verbs API, enabling interconnect portability~\cite{nvidia2022rdmacomparison}.

For GPU-initiated communication, the hardware requirement is NIC support for
device-accessible control structures. NVIDIA ConnectX series adapters
(ConnectX-6~Dx and later) and BlueField DPUs provide this capability through DOCA
GPUNetIO~\cite{doca-gpunetio}. Systems lacking such hardware support cannot
enable direct GPU-NIC communication and must fall back to CPU-mediated
mechanisms. In GIN's architecture, this constraint motivates the dual-backend
design: the GDAKI backend leverages DOCA GPUNetIO for direct device
communication on supported hardware, while the Proxy backend provides
functionally equivalent semantics through lock-free GPU-to-CPU queues and
CPU-driven network operations on arbitrary RDMA-capable NICs.


Optimal performance requires co-location of GPUs and NICs on the same PCIe root
complex to minimize peer-to-peer latency and maximize
bandwidth~\cite{nvidia2013gpudirect,nvidia2021gpudirectreqs}. Multi-socket
systems with distributed PCIe topologies may incur inter-socket traversal
penalties, reducing GPUDirect RDMA efficiency. Additionally, GPU-initiated
networking requires the \texttt{nv\_peer\_mem} kernel module (for GPUDirect
RDMA) and appropriate driver stacks (OFED for InfiniBand, MOFED for Mellanox
adapters) to establish memory mappings between GPU and NIC address
spaces~\cite{nvidia2021gpudirectreqs}.


\subsection{NCCL Architecture and Network Plugins}

\textbf{NCCL} is the standard collective communication runtime for multi-GPU
machine learning, providing topology-aware implementations of allreduce,
allgather, reduce-scatter, and broadcast operations~\cite{hu2025demystifying}.
NCCL's architecture uses CPU proxy threads to orchestrate network operations,
where GPU kernels enqueue communication descriptors into host-visible queues and
CPU threads execute them via network plugins. While this design has proven
robust for collective operations at scale, NCCL 2.28's Device API~\cite{nccl2.28} extends this 
architecture with device-side primitives that enable applications to implement 
custom communication patterns directly from GPU code, integrate communication 
within compute kernels, and achieve fine-grained computation-communication overlap 
for emerging workloads.

The widespread adoption of NCCL in production frameworks motivated this integration
of device-initiated communication capabilities, preserving ecosystem compatibility 
while enabling new use cases.

\textbf{NCCL Network Plugin} architecture provides an abstraction layer that decouples the core 
library from specific network implementations. NCCL supports both internal plugins 
(e.g., Socket and InfiniBand) built directly into the library, and external plugins 
that implement the NCCL network API as shared libraries (\texttt{libnccl-net.so}). 
This design allows network vendors and hardware providers to extend NCCL with 
specialized transport implementations without modifying the NCCL core. External 
plugins are dynamically loaded at runtime, selected via the \texttt{NCCL\_NET\_PLUGIN} 
environment variable, enabling seamless integration of diverse networking technologies 
while maintaining version compatibility through a versioned API interface. 

\subsection{Specialized MoE Communication Libraries}

MoE architectures in LLMs require dynamic,
load-balanced all-to-all token routing with unpredictable message sizes,
creating irregular communication patterns 
that traditional collectives~\cite{ren2024deepseekmoe} are not well suited for. 
Specialized libraries like
\textbf{DeepEP}~\cite{deepep2025} and \textbf{Perplexity's pplx-kernels}~\cite{pplx-kernels}
target these workloads with CUDA-optimized, GPU-initiated primitives for
low-latency all-to-all transfers. These efforts demonstrate the value of
GPU-centric communication for MoE workloads but remain separate from NCCL's
ecosystem.

GIN brings GPU-initiated RDMA operations into NCCL, enabling applications to
leverage both host-optimized collectives and device-driven point-to-point
communication within a unified runtime.

Building on these device-initiated networking technologies and NCCL's plugin 
architecture, the next section presents GIN's design and implementation.

\section{NCCL GIN: GPU-Initiated Networking}
\label{sec:nccl-gin}

This section presents the design and implementation of GIN for NCCL. As 
established in Section~\ref{sec:background}, extending NCCL with device-side 
primitives enables modern AI workloads to achieve tight coupling between 
computation and communication. GIN provides this capability by integrating 
device-initiated one-sided primitives into NCCL, allowing GPU threads to 
initiate network operations directly from CUDA kernels without CPU involvement.

The design preserves NCCL's established programming model and ecosystem integration while adding a parallel low-latency path for device-driven communication. This enables production systems like TensorRT-LLM, vLLM, and SGLang, as well as communication libraries like DeepEP, to implement custom collective algorithms and kernel fusion patterns previously unattainable with conventional NCCL.

We organize the section in three parts: Section~\ref{sec:ncclgin-principles} introduces the three-layer architecture and core design principles; Section~\ref{sec:ncclgin-api} presents the device-side API and demonstrates its usage through a practical example; and Section~\ref{sec:ncclgin-backends} analyzes the two plugin interfaces and their backend implementations, GDAKI and Proxy, explaining their design rationale and performance characteristics.

\subsection{Core Principles and Architecture}
\label{sec:ncclgin-principles}

NCCL GIN's architecture is built on one-sided communication semantics,
which enable two critical performance objectives: \emph{maximum communication-computation 
overlap} and \emph{minimum end-to-end operation latency}. By enabling GPU threads 
to initiate RDMA operations directly---reading from and writing to remote 
memory without receiver coordination---GIN eliminates host-device synchronization 
overhead and two-sided handshaking delays, allowing asynchronous data transfers to 
proceed concurrently with computation.

\begin{figure}[t]
\centering
\includegraphics[width=0.95\columnwidth]{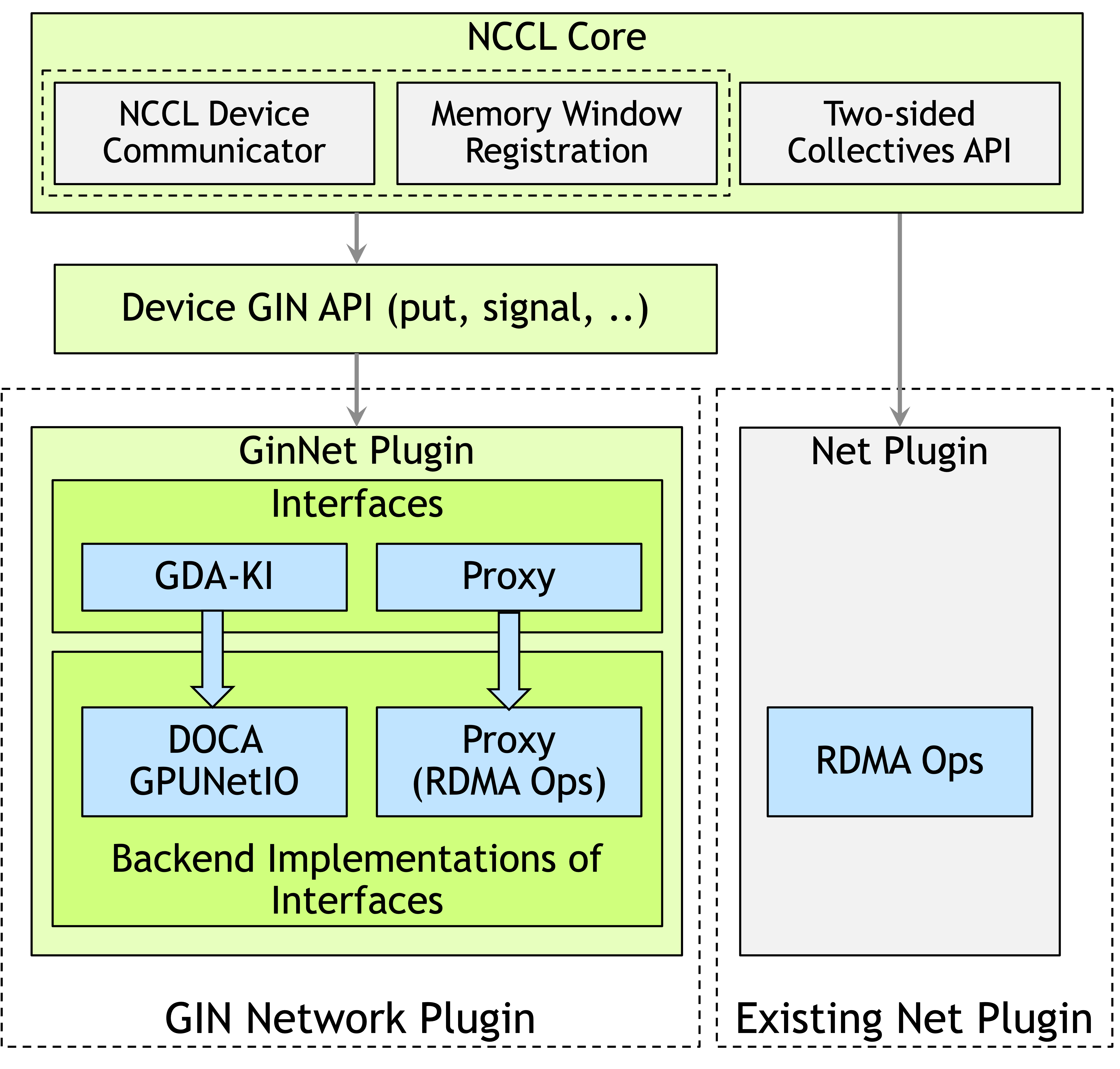}
\caption{GIN Architecture showing the interactions between NCCL Core, Plugin Layer, and Device-Side API.}
\label{fig:gin-arch}
\end{figure}

The architecture comprises three cooperating layers for GIN---shown in the green path (left side) of
Figure~\ref{fig:gin-arch}---designed to balance high performance with broad vendor
support: \emph{NCCL Core} (host-side APIs), \emph{Device GIN API} (GPU-callable 
primitives), and \emph{GIN Network Plugin} (pluggable network backends). The grey path (right side) of
Figure~\ref{fig:gin-arch} shows the existing two-sided collectives API over the built-in network plugin.
We discuss the three GIN layers further:
\textbf{i)~NCCL Core}: the host-side NCCL functionality manages 
memory-window registration, resource allocation, and communicator initialization, 
providing the foundation for GIN resource management and extending NCCL's existing 
infrastructure with device-initiated communication capabilities;
\textbf{ii)~Device GIN API}: the device-side API exposes a unified interface 
to GPU kernels, enabling applications to invoke one-sided communication operations 
directly from CUDA kernels. It dispatches to either NCCL-provided (Proxy) or 
plugin-provided (GDAKI) implementations based on the underlying network backend;
\textbf{iii)~GIN Network Plugin}: the plugin layer provides an 
extensibility mechanism that defines remote data-movement operations and supports 
dual semantics—GDAKI and Proxy—to maximize network coverage. NCCL's InfiniBand 
transport implements both, while external vendors may supply their own. Under the 
Proxy interface, NCCL Core owns control structures, device-side queuing logic, and 
device API implementations, while plugins provide only CPU-based \texttt{put}, 
\texttt{signal}, \texttt{test}, and \texttt{regMr} operations, enabling networks 
without GPU-direct capabilities and lowering the barrier to GIN adoption. Under 
GDAKI semantics, plugins own both the control path and device API: they create 
GPU contexts via \texttt{createContext} and supply device code that directly 
programs NICs using kernel-initiated APIs such as DOCA GPUNetIO, with NCCL Core 
coordinating the structure exchange between host and device components.

These components enable device-initiated one-sided communication through 
several key design elements: \emph{one-sided semantics} for unilateral data 
movement, \emph{symmetric memory windows} for zero-copy remote access, and 
\emph{asynchronous completion tracking} with flexible ordering semantics.

\noindent\textbf{One-Sided Communication Semantics.}
GIN exposes one-sided RDMA primitives---\texttt{put} for remote writes 
and \texttt{put} with \texttt{signal} for writes with remote 
notification---enabling GPU threads to access remote memory without any receiver 
coordination. This one-sided model eliminates the overhead of handshaking protocols and the need for receiver 
participation, allowing initiators to issue transfers unilaterally and 
independently control when to verify completion. 
The one-sided model proves particularly effective for irregular communication 
patterns in MoE workloads, where dynamic token routing creates unpredictable 
traffic patterns, and for single-shot collective implementations 
that benefit from parallel, non-blocking peer communication.

\noindent\textbf{Windows-based~\textit{(A)}Symmetric Memory.}
Communication buffers must be collectively registered across all ranks, 
establishing memory windows that are symmetric in addressability, following 
the MPI RMA window model~\cite{10.1145/2780584}. All processes 
can access registered memory, analogous to NVSHMEM's symmetric heap. GIN windows are designed to support 
asymmetry in capacity: each rank may register different buffer sizes. This flexibility proves essential 
for disaggregated serving architectures where prefill ranks require larger 
buffers than decode ranks. Note that the current implementation in NCCL 2.28 
enforces symmetric sizes, though this constraint will be lifted in future releases. 
Furthermore, the memory allocation is not coupled with the registration: 
NCCL-GIN memory windows allow users to create a window from an existing allocation. 
Each registration produces window handles 
encapsulating remote access metadata. The window handles provide backend-specific optimization opportunities.
Backends construct RDMA descriptors directly from window metadata and target addresses using rank-relative offsets.

\noindent\textbf{GIN Contexts for Network Parallelism.}
GIN contexts serve as the primary abstraction for expressing network parallelism.
Each context abstracts a channel between the GPU and NIC and encapsulates network resources and connections (queue pairs (QPs)). 
Multiple contexts per communicator enable applications to exploit network-level parallelism across multiple NICs, ports, and QPs, allowing independent concurrent communication streams. A single context can address every rank associated with the communicator. As such, a context can issue multiple 
concurrent operations to different peers. 

\noindent\textbf{Asynchronous Completion Tracking.}
All device-initiated operations execute asynchronously and return immediately 
to enable other work to proceed in parallel, such as computation or intra-node 
communication through NVLink. Applications track operation 
completion through two distinct mechanisms using per-context resources. 
\emph{Counters} are local objects that track completion on the sender side, indicating when source buffers can be safely reused. Unlike the \emph{flush} operation that tracks the completion of all operations posted to the context, \emph{Counters} are a powerful concept to track local completions per operation, allowing users to describe pipeline algorithms efficiently. Each data movement operation can optionally report local completion to a counter provided by the user (via counterID). 

On the other hand, \emph{Signals}, which are symmetric objects, provide remote completion tracking, confirming 
data arrival and visibility at the destination. 
Unlike OpenSHMEM's address-based synchronization, GIN uses ID-based addressing: 
each signal (and counter) is identified by an integer ID rather than a memory 
address. This ID-based design simplifies resource management and enables 
efficient hardware implementation of completion notifications.

\noindent\textbf{Ordering Semantics.}
To maximize network efficiency and throughput, GIN operations are unordered by default. However, 
GIN provides ordering guarantees 
only between \texttt{put} and \texttt{signal} operations in the same context to the same peer. 
When a \texttt{signal} operation 
(either standalone \texttt{signal} or \texttt{put} with \texttt{signal} 
action \texttt{ncclGin\_SignalInc}/\texttt{ncclGin\_SignalAdd}) 
completes at the destination, it guarantees that all preceding \texttt{put} operations 
to that peer on the same context have completed and are visible to remote GPU threads. 
This provides lightweight ordering without requiring explicit fence operations: 
applications can batch multiple \texttt{put}s and attach a \texttt{signal} to the final operation, 
ensuring ordered remote visibility of the entire sequence. In contrast, the \texttt{flush} operation 
ensures only local completion---all pending operations have been consumed and source 
buffers can be safely reused---but makes no guarantees about remote visibility. 
GIN's signal-based ordering aims for maximum performance 
by providing ordering guarantees selectively through signals rather than globally.
It is worth noting that GIN does not assume or guarantee any ordering between GPU threads. It is the responsibility of the user to synchronize the threads using CUDA synchronization primitives. 
\subsection{Device-Side API and Programming Model}
\label{sec:ncclgin-api}

The device-side API provides GPU kernels with direct control over network operations, exposing methods invokable from CUDA device code without CPU intervention. The programming model centers on the \texttt{ncclGin} object, which encapsulates networking resources (contexts, operation queues, and peer connection state) and provides methods for data movement, completion tracking, and synchronization. Backend selection (DOCA GPUNetIO or Proxy) occurs transparently at communicator initialization based on hardware capabilities and user configuration, presenting an identical device-facing interface regardless of the underlying implementation.

\noindent\textbf{API Organization.}
The interface organizes operations into four logical categories that reflect the communication workflow. \emph{Data movement} operations (\texttt{put}, \texttt{putValue}, \texttt{signal}) initiate one-sided transfers or notifications to remote peers, posting RDMA operations that execute asynchronously. \emph{Completion tracking} operations distinguish between local completion (\texttt{flush}, \texttt{readCounter}, \texttt{waitCounter})—which signals that source buffers can be safely reused—and remote completion (\texttt{readSignal}, \texttt{waitSignal})—which confirms data arrival at the destination and visibility to remote GPU threads. \emph{Barrier synchronization} (\texttt{ncclGinBarrierSession}) coordinates all ranks within a team before communication phases, ensuring global consistency through network-wide synchronization. \emph{State management} operations (\texttt{resetCounter}, \texttt{resetSignal}) reset completion state for reuse across multiple communication rounds. This separation of concerns enables fine-grained control over communication-computation overlap and supports diverse synchronization patterns.

Furthermore, GIN device API has a tight interaction with the GPU memory model and takes hints from users to optimize performance-critical memory ordering and consistency tasks. For instance, \texttt{put} takes two hints from the user on the provided visibility scope of the data as well as the expected user visibility after the operation completes.   

\noindent\textbf{Usage Workflow.}
Applications interact with GIN through a three-phase workflow 
(Listing~\ref{lst:ncclgin-api}). During initialization, applications enable GIN 
support when creating the NCCL Device communicator (\texttt{ncclDevCommCreate} 
with appropriate configuration flags), which creates the GIN contexts. 
Applications then collectively register memory buffers as windows using 
\texttt{ncclCommWindowRegister}, which returns window handles for use in device 
code. During kernel execution, device threads instantiate an \texttt{ncclGin} 
object specifying the desired context index (typically selected based on target 
peer or load balancing requirements) and issue data movement operations, 
optionally attaching completion actions such as remote signal increments or 
local counter updates. Finally, kernels synchronize by waiting on signals or 
counters before reusing buffers or advancing to subsequent computation stages, 
ensuring proper ordering of communication and computation phases. 
Listing~\ref{lst:ncclgin-api} presents a simplified view of the NCCL GIN 
interface, highlighting the essential operations. The actual API includes 
additional template parameters and options for advanced use cases (e.g., 
cooperative thread groups, inline data transfers), but the core abstraction 
remains consistent.

\begin{figure}[ht]
\begin{lstlisting}[caption={Simplified NCCL GIN device API.}, label={lst:ncclgin-api}]
class ncclGin {
  // Constructor: initialize with device
  // communicator and context ID
  ncclGin(ncclDevComm comm, int contextIndex);

  // Data Movement Operations
  void put(team, peer, dstWindow, dstOffset, 
           srcWindow, srcOffset, bytes, ...);
  void putValue(team, peer, dstWindow, dstOffset, value, ...);
  void signal(team, peer, signalId);

  // Local Completion Tracking
  void flush(coop); // Block until ops complete
  uint64_t readCounter(counterId); // Poll counter
  void waitCounter(coop, counterId, expectedValue);
  void resetCounter(counterId); // Reset for reuse

  // Remote Completion Tracking
  uint64_t readSignal(signalId); // Poll signal
  void waitSignal(coop, signalId, expectedValue);
  void resetSignal(signalId); // Reset for reuse
};

// Network Barrier for cross-rank synchronization
class ncclGinBarrierSession {
  ncclGinBarrierSession(coop, gin, team, barrierHandle, index);
  void sync(coop); // Global barrier sync
};

// Optional: Attach completion actions to data
// Remote signal
put(..., ncclGin_SignalInc{signalId});
// Local counter
put(..., ncclGin_CounterInc{counterId});
\end{lstlisting}
\end{figure}

\noindent\textbf{Usage Example.}
Listing~\ref{lst:ring-exchange} demonstrates a unidirectional ring exchange pattern using GIN primitives. In this kernel, each rank sends data to its successor in a ring topology (\texttt{myRank + 1}), with data flowing in one direction around the ring, implementing a common pattern in pipelined communication algorithms. The \texttt{put} operation (lines~13--16) transfers data from the local \texttt{sendWin} to the peer's \texttt{recvWin} at a computed offset and atomically increments remote signal~0 upon completion, providing remote notification of data arrival. The sending rank then waits for its own signal~0 to be incremented by its predecessor (line~19), ensuring that received data has arrived and is visible to local GPU threads before proceeding with computation. Finally, the signal is reset for subsequent communication rounds (line~21). This pattern illustrates how GIN's asynchronous operations, flexible completion semantics, and explicit synchronization primitives enable efficient overlapped point-to-point communication from device code.

\begin{figure}[ht]
\begin{lstlisting}[caption={Unidirectional ring exchange using NCCL GIN.},label={lst:ring-exchange},numbers=left,numberstyle=\tiny,stepnumber=1,basicstyle=\footnotesize\ttfamily,xleftmargin=2em,framexleftmargin=1.5em]
__global__ void ringExchange(
  ncclDevComm devComm, 
  ncclWindow_t sendWin, 
  ncclWindow_t recvWin,
  size_t dataSize, int myRank)
{
  // Initialize ctx 0
  ncclGin gin(devComm, 0);
  int peer = (myRank + 1) % devComm.nRanks;

  // Send data to peer and signal completion
  // by incrementing peer's signal
  gin.put(ncclTeamWorld(devComm), peer,
           recvWin, myRank * dataSize,
           sendWin, peer * dataSize, dataSize,
           ncclGin_SignalInc{0} );

  // Wait for predecessor
  gin.waitSignal(ncclCoopCta(), 0, 1);
  // Reset for next round
  gin.resetSignal(0); 
}
\end{lstlisting}
\end{figure}

\subsection{Backend (GDAKI and Proxy) Implementations}
\label{sec:ncclgin-backends}

The device API abstracts two distinct backend implementations that realize the 
dual plugin semantics described earlier. The GDAKI backend implements the 
GDAKI semantics through direct GPU-to-NIC communication via DOCA GPUNetIO, 
with the plugin providing both device API and control path. The Proxy backend 
implements Proxy semantics through CPU-mediated transfers, with NCCL Core 
providing the device API and the plugin providing only CPU-based data path 
operations. Both backends expose an identical device-facing interface, enabling 
transparent backend selection at runtime without application code changes.

\noindent\textbf{GDAKI Backend: Direct GPU-to-NIC Communication.}
The GDAKI backend implements device-initiated networking in its purest form by 
leveraging DOCA GPUNetIO to enable GPU threads to directly program network 
interface cards without CPU involvement. When a kernel invokes \texttt{put}, 
GPU threads construct RDMA work queue entries (WQEs) in device memory, populate 
them with source/destination addresses and transfer metadata, and directly write 
to the NIC's doorbell registers to trigger DMA transfers. The NIC hardware 
manages operation progress autonomously: it polls GPU memory for new WQEs, 
executes RDMA transactions over InfiniBand or RoCE, and updates completion 
queue entries in GPU-visible memory. This direct GPU-NIC path eliminates PCIe 
round-trips to the CPU and achieves low-latency for small 
messages. However, the approach requires modern hardware and software: 
ConnectX-6~Dx or newer NICs with GPU-accessible control structures and 
required CUDA 12.x version (as detailed in~\cite{doca-gpunetio}).
Additionally, 
proper system configuration---including GPUDirect RDMA kernel modules 
(\texttt{nv\_peer\_mem} or \texttt{dmabuf}) and co-located GPU-NIC PCIe topology---is essential 
for correct operation and optimal performance.

\noindent\textbf{Proxy Backend: CPU-Assisted Communication.}
The Proxy backend trades peak performance for hardware portability by 
introducing the CPU into the communication path as an intermediary between the GPU 
and NIC. GPU threads enqueue operation descriptors---64-bytes containing source/destination 
window handles, potential source inline value, offsets, sizes, and completion actions---into lock-free queues 
allocated in CPU memory using fire-and-forget stores. A dedicated CPU proxy thread per communicator, 
pinned to NUMA nodes near the local rank's GPU and NIC, continuously polls these queues. 
Upon detecting new descriptors, the proxy thread extracts the fields and posts the network operation through 
the network plugin's \texttt{iput}/\texttt{iput\_signal} interface, which maps to standard 
InfiniBand verbs or other network APIs. The plugin is responsible for executing the 
signal and ensuring visibility of all prior \textit{put} operations. Completion notifications follow the 
reverse path: the proxy thread polls on completions using the network plugin's \textit{test} interface, 
matches completed operations to their associated GIN counters, and updates 
completion state in GPU-visible memory (GPU or CPU-resident based on GDRCopy availability). 
While this CPU involvement introduces additional latency compared to GDAKI, the Proxy backend supports 
arbitrary CUDA versions, any GPUDirect RDMA-capable NIC (InfiniBand, RoCE, iWARP), and Volta-or-newer GPUs. 
Additionally, CPU involvement simplifies debugging through host-side 
instrumentation and enables graceful performance degradation on systems where direct 
GPU-NIC communication is unavailable.

\noindent\textbf{Backend Selection and Portability.}
Table~\ref{tab:backend-comparison} summarizes the architectural differences between GDAKI and Proxy backends. High-performance production systems with modern NVIDIA end-to-end networking infrastructure (ConnectX-6~Dx or newer NICs, recent CUDA versions, properly configured GPUDirect RDMA) favor GDAKI for minimal latency and zero CPU overhead. Development environments, legacy hardware deployments, multi-vendor network fabrics, or systems with misconfigured GPUDirect support rely on Proxy for functional correctness and operational flexibility. The runtime automatically detects available backends during communicator initialization (\texttt{ncclCommInitRank}), probing for DOCA GPUNetIO support through capability queries and falling back to Proxy when necessary. Applications can override this selection through environment variables (\texttt{NCCL\_GIN\_BACKEND}) for debugging or performance tuning. This design ensures portability across diverse deployment scenarios while preserving the performance benefits of direct GPU-NIC communication where hardware and software infrastructure permit.

\begin{table}[t]
\footnotesize
\centering
\caption{Architectural comparison of NCCL GIN backend implementations}
\label{tab:backend-comparison}
\begin{adjustbox}{max width=\columnwidth}
\begin{tabular}{@{}p{2.3cm}p{2.6cm}p{2.4cm}@{}}
\toprule
\textbf{Characteristic} & \textbf{GDAKI} & \textbf{Proxy} \\
\midrule
\textbf{Comm. Path} & Direct GPU$\leftrightarrow$NIC & GPU$\rightarrow$CPU$\leftrightarrow$NIC \\
\addlinespace[0.5em]
\textbf{CPU Involvement} & Zero (fully device-driven) & Required (dedicated thread per comm) \\
\addlinespace[0.5em]
\textbf{Progress Model} & NIC hardware autonomously polls GPU memory & CPU thread polls queues and posts to NIC \\
\addlinespace[0.5em]
\textbf{Operation Posting} & GPU directly rings NIC doorbell & GPU writes descriptor; CPU extracts and posts \\
\addlinespace[0.5em]
\textbf{Hardware Requirements} & \parbox{2.6cm}{\vspace{1mm}ConnectX-6 Dx+ NIC\\CUDA 12.2+\vspace{1mm}} & \parbox{2.4cm}{\vspace{1mm}Any RDMA NIC\\Any CUDA GPU\\Any CUDA version\vspace{1mm}} \\
\addlinespace[0.5em]
\textbf{Implementation Base} & DOCA GPUNetIO device verbs & Plugin \texttt{iput}/\texttt{test} API \\
\addlinespace[0.5em]
\textbf{Debugging Support} & Device-side tools only & Host-side inspection and tracing \\
\addlinespace[0.5em]
\textbf{Portability} & Requires GPU-NIC direct access (reference impl: ConnectX) & Universal (all vendors) \\
\addlinespace[0.5em]
\textbf{Use Case} & Production HPC/AI clusters & Development, legacy, multi-vendor \\
\bottomrule
\end{tabular}
\end{adjustbox}
\end{table}

\section{DeepEP Integration}
\label{sec:integration}

This section presents NCCL GIN integration into DeepEP to validate its effectiveness
for workloads requiring computation-communication fusion and low-latency.
DeepEP is a specialized MoE communication library that implements device-initiated sparse
all-to-all communication---dispatch and combine primitives---using NVSHMEM and IBGDA. 
The library provides two flavors of dispatch and combine primitives: high-throughput
(HT) and low-latency (LL) kernels used in training/inference-prefill
and inference-decode phases, respectively. This integration
demonstrates how DeepEP's device-initiated communication patterns 
can be implemented using GIN APIs while preserving performance characteristics 
and maintaining coexistence with the existing NVSHMEM communication backend.

\subsection{Integration Requirements}

DeepEP's communication patterns impose several requirements: \textbf{i)~High QP
Parallelism}---HT kernels require 24 QPs, while LL kernels require 8-16
QPs matching local expert count; \textbf{ii)~Heterogeneous Topology
Support}---HT uses symmetric rank-to-rank RDMA with NVLink forwarding, while LL
uses full all-to-all RDMA mesh; \textbf{iii)~Fine-Grained Synchron\-ization}---atomic
updates to head/tail pointers for circular buffer flow control; \textbf{iv)~Backend 
Coexistence}---NVSHMEM IBGDA and GIN backends must coexist to match user preference 
according to the execution environment. 

\subsection{Backend Integration Strategy}

The integration employs a minimal abstraction layer for lifecycle management
(initialization, memory allocation, barriers) while allowing performance-critical
operations to use backend-specific device APIs directly in kernels via
conditional compilation. This design accommodates fundamental semantic
differences: IBGDA uses pointer-based addressing with memory atomics, while NCCL
GIN uses window-based addressing with signal atomics.

The integration addresses four key translation challenges. \textbf{First},
multi-communicator mapping: since NCCL GIN provides 4 contexts per communicator,
meeting DeepEP's QP requirements requires
$\lceil\text{QPs}/4\rceil$ communicators, with work distributed via
deterministic selection (\texttt{comm\_id = id / 4}, \texttt{ctx\_id = id \%
4}). \textbf{Second}, memory management: the backend registers allocated buffers
with all communicators and stores device-accessible window handles in GPU
memory, enabling kernels to translate pointer arithmetic to (window, offset)
pairs.
\textbf{Third}, synchronization: pre-allocated structured signal layouts map
memory-based atomics to signal primitives (HT: two signals per channel for
head/tail; LL: one signal per expert). \textbf{Fourth}, semantic preservation:
zero-byte \texttt{put}s with atomic signals emulate release-acquire semantics, ensuring
visibility of prior transfers before signaling completion.

\begin{table*}[t]
\centering
\footnotesize
\begin{tabular}{@{}p{2.5cm}p{5.8cm}p{5.8cm}@{}}
\toprule
\textbf{Aspect} & \textbf{DeepEP-Custom NVSHMEM/IBGDA Layer} & \textbf{NCCL GIN} \\
\midrule
\textbf{Memory Model} & 
\emph{PGAS}: Partitioned Global Address Space with symmetric heap; direct pointer-based addressing across PEs &
\emph{Window-based}: Explicit registration of memory windows; offset-based addressing within windows \\
\addlinespace[0.5em]

\textbf{Data Transfer API} & 
\texttt{put\_nbi(dst\_ptr, src\_ptr, count, pe)} --- pointer-based non-blocking \texttt{put}s; warp-collective execution; asynchronous one-sided &
\texttt{put(team, peer, dstWin, dstOff, srcWin, srcOff, bytes)} --- window/offset pairs; thread-based; asynchronous one-sided \\
\addlinespace[0.5em]

\textbf{Synchronization Primitives} & 
\emph{Memory atomics}: Direct atomic operations (\texttt{atomic\_add}, \texttt{atomic\_fetch}) on remote memory locations &
\emph{Signal atomics}: Dedicated signal infrastructure; \texttt{signal(peer, id)} for atomic updates, \texttt{readSignal(id)} for polling \\
\addlinespace[0.5em]

\textbf{Completion Model} & 
\emph{Local flush}: \texttt{quiet()} per QP; blocks until all operations complete on a particular QP for a unique remote PE&
\emph{Per-context flush}: \texttt{flush()} per context enables parallel completion checking across multiple queues \\
\addlinespace[0.5em]

\textbf{Barrier Operations} & 
Team-based barriers (\texttt{barrier(team)}, \texttt{barrier\_all()}) with opaque team handles; synchronizes all team members &
\texttt{ncclGinBarrierSession} with team tags and session IDs; enables symmetric subset synchronization without full team coordination \\
\bottomrule
\end{tabular}
\caption{\small Selective DeepEP-Custom NVSHMEM/IBGDA and NCCL GIN APIs used by DeepEP Communication Kernels.\\\textbf{NOTE:} This table is not a comparison of NVSHMEM/IBGDA and NCCL GIN APIs---it focuses on corresponding NVSHMEM and NCCL APIs used in DeepEP library.}
\label{tab:operation-mapping}
\end{table*}

\subsection{Operation Semantic Mapping}

The migration from NVSHMEM to NCCL GIN requires translating between distinct
programming models while preserving communication semantics. NVSHMEM provides a
PGAS (Partitioned Global Address Space) abstraction with pointer-based
addressing and memory-based synchronization primitives, while NCCL GIN follows a
window-based one-sided model with signal-based completion tracking
(reviewed earlier in Section~\ref{sec:ncclgin-api}). Table~\ref{tab:operation-mapping} maps DeepEP's
communication patterns to the corresponding NVSHMEM and GIN primitives, highlighting the key
semantic transformations required for integration.
For data transfers, kernels compute
window-relative offsets on-the-fly and
select communicators deterministically based on channel or expert ID, load balancing
across QPs. For synchronization, the signal-based design decouples
data movement from completion notification: bulk transfers use \texttt{put()}
without immediate signaling, followed by explicit \texttt{signal()} operations
that atomically update remote counters only after all prior operations complete.
This pattern implements release-acquire semantics through network primitives
rather than memory ordering, ensuring data visibility at the completion of signal arrival.

\subsection{High-Throughput Kernel Integration}

HT kernels optimize for large batches (4096 tokens) using hierarchical
communication. GPUs send data to remote nodes over symmetric RDMA connections,
which then forward tokens via NVLink to destination GPUs. This minimizes
inter-node traffic while maximizing intra-node bandwidth. The RDMA buffer
contains multiple channels functioning as QPs, each with send/receive buffers.
Head and tail pointers track buffer occupancy, providing circular-buffer flow
control.

The dispatch kernel assigns specialized roles to SMs. Odd-numbered SMs act as
Senders (transmitting tokens to remote ranks) and NVLink Receivers (final
destinations), while even-numbered SMs act as Forwarders (receiving RDMA tokens
and forwarding them via NVLink). This specialization enables concurrent,
bidirectional communication. To reduce contention, separate channels are used
for data/tail updates and for head-pointer updates, distributing work across
distinct communicators.

Following the operation mappings (Table~\ref{tab:operation-mapping}), each SM
role uses signal-based atomics for pointer management and window-based
\texttt{put()} for data transfers. Remote tail signals are incremented using
\texttt{gin.signal(SignalAdd, 1)}, while head-pointer flow control relies on
polling local head signals via \texttt{gin.readSignal(signal\_id)}. Data
transfers use single-threaded NCCL GIN \texttt{put()} followed by
\texttt{\_\_syncwarp()} to preserve warp-collective semantics. The notify
dispatch kernel uses a coordinator SM to flush all writes
(\texttt{gin.flush()}), perform a barrier across symmetric RDMA ranks, reset
head/tail signals, and exchange metadata before starting the main dispatch.

The combine kernel mirrors the dispatch kernel’s specialization, using 25
warps per SM. Even-numbered SMs serve as NVLink Senders (distributing input
tokens to local buffers), RDMA Receivers (integrating remote tokens with bias
terms), and Coordinators monitoring receiver progress. Odd-numbered SMs act as
NVLink and RDMA Forwarders (merging local tokens and forwarding them to remote
ranks), with corresponding Coordinators monitoring forwarders. The operation
mapping parallels dispatch: Forwarder warps use single-threaded \texttt{put()}
with \texttt{\_\_syncwarp()} for data transfer, \texttt{readSignal()} for
head-pointer polling, and \texttt{signal()} for tail updates. Receiver warps
use \texttt{readSignal()} for tail-pointer monitoring, while Coordinator warps
update head pointers via \texttt{signal()}.

\subsection{Low-Latency Kernel Integration}

LL kernels optimize for small batches (1-128 tokens) using full all-to-all RDMA
mesh connectivity, enabling direct GPU-to-GPU communication. Token streaming
embeds routing metadata without separate notify phases, minimizing
dispatch-combine cycle time. Per-expert signal allocation provides direct
coordination between any expert pair across the cluster.

SM allocation distributes experts using $G = \lceil N / S \rceil$ warp groups
per SM, where $N$ is total experts and $S$ is available SMs. Each SM is assigned
experts via \texttt{expert\_idx = sm\_id * G + warp\_group\_id}. Within each
warp group, most warps handle FP8 quantization and token sending, while a
counting warp manages expert counts and metadata for all assigned experts. This
organization enables efficient parallelization across hundreds of experts (e.g.,
288 experts across 132 SMs with 3 warp groups per SM).

LL kernels leverage hybrid NVLink-RDMA communication. For each token transfer,
kernels check NVLink availability via custom function \texttt{nccl\_get\_p2p\_ptr}. If
available, tokens are copied directly using warp-level memory operations;
otherwise, NCCL GIN's \texttt{put()} performs RDMA transfers
(Table~\ref{tab:operation-mapping}). Tokens are first copied from PyTorch
tensors into RDMA send buffers, with optional FP8 quantization applied during
this stage. After completing token transfers to a destination, a counting warp
sends the per-expert token count using zero-byte \texttt{put()} with
\texttt{SignalAdd}, ensuring all prior data transfers have completed and become
visible before the count is delivered---this implements the release-acquire
semantics described in the Backend Integration Strategy. Receivers poll using
\texttt{gin.readSignal(signal\_id)} until tokens arrive.


The combine kernel routes expert outputs back to source ranks with weighted
reduction. It uses the same hybrid NVLink-RDMA approach with optional LogFMT
compression to reduce data volume. 
After transmitting expert outputs, flag
signals notify destinations using zero-byte \texttt{put()} combined with
\texttt{SignalAdd}, ensuring completion before receivers begin accumulation.
Receivers employ TMA load warps to fetch expert outputs into shared memory,
then apply top-k weights using reduction warps in FP32 before converting to
BF16 output.

\section{Performance Evaluation}
\label{sec:results}

This section evaluates NCCL GIN and NVSHMEM through two complementary approaches. 
We begin with point-to-point microbenchmarks (Section~\ref{sec:results:microbenches}) 
to isolate protocol-level performance characteristics, then assess integration with 
DeepEP~version~1.2.1, a production MoE communication library. The DeepEP evaluation\footnote{While 
NVSHMEM supports both IBGDA and IBRC transports, DeepEP's implementation is tightly coupled with IBGDA.} 
covers High-Throughput (HT) kernels for training and inference-prefill (Section~\ref{sec:results:ht}), 
and Low-Latency (LL) kernels for inference-decode (Section~\ref{sec:results:ll}) under both 
hybrid RDMA+NVLink and pure RDMA configurations.

All experiments run on NVIDIA's EOS cluster with H100 GPUs (Table~\ref{tab:specs}), 
using NVSHMEM version~3.4.5 and NCCL version~2.28. DeepEP benchmarks allocate 24 SMs per GPU, 
with communication distributed across NVLink and RDMA based on automatically 
selected channel configurations.

\begin{table}[t]
\centering
\caption{Hardware specification of EOS DGXH100 compute nodes.}
\label{tab:specs}
\footnotesize
\begin{tabularx}{0.85\columnwidth}{@{}XX@{}}
\toprule
\textbf{Specification} & \textbf{DGXH100 Node} \\
\midrule
Number of Nodes     & 576 \\
GPU Model           & H100 80GB HBM3 \\
GPUs per Node       & 8 (640 GB total) \\
GPU Memory BW       & 3.2 TB/s \\
NVLink Generation   & 4th Generation \\
NVLink BW           & 900 GB/s bidirectional \\
NVLink per GPU      & 18 links \\
CPU Model           & Intel Xeon Platinum 8480CL \\
CPU Sockets         & 2 \\
CPU Cores           & 112 (56 per socket) \\
CPU Clock           & 2.0 GHz base, 3.8 GHz boost \\
System Memory       & 2 TB \\
InfiniBand          & 8$\times$400 Gbit/s (compute) \\
                    & 2$\times$400 Gbit/s (storage) \\
\bottomrule
\end{tabularx}
\end{table}

\subsection{Point-to-Point Microbenchmarks}
\label{sec:results:microbenches}

\begin{figure}[t]
    \centering
    \includegraphics[width=\columnwidth]{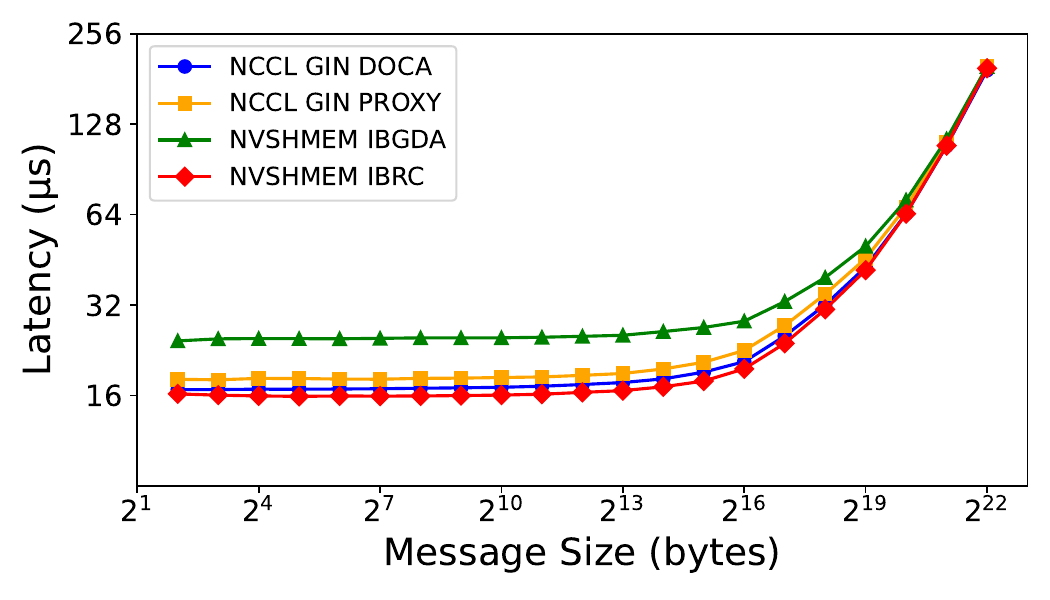}
    \caption{Point-to-point latency: NVSHMEM IBGDA/IBRC and NCCL GIN GDAKI/Proxy backends.}
    \label{fig:results:microbenches}
\end{figure}

To establish baseline performance, we measure \texttt{put} with signal latency using ping-pong 
tests across message sizes from 4~bytes to 4~MB between two H100 GPUs. Figure~\ref{fig:results:microbenches} 
presents the performance of NCCL GIN's dual backends (GDAKI and Proxy) 
alongside NVSHMEM IBGDA and IBRC transports.

For small messages (4--128 bytes), NCCL GIN GDAKI achieves 16.7~$\mu$s round-trip latency, 
comparable to NVSHMEM IBRC at 16.0~$\mu$s, while NVSHMEM IBGDA achieves 24.3~$\mu$s. 
The GDAKI backend's direct GPU-to-NIC path eliminates CPU proxy overhead, while the Proxy 
backend achieves 18.0~$\mu$s despite GPU-to-CPU queue traversal. At larger message 
sizes, bandwidth limitations dominate and all implementations converge, validating NCCL GIN's 
fundamental performance characteristics for application integration.

\subsection{High-Throughput Kernels}
\label{sec:results:ht}

\begin{figure}[t]
    \centering
    \includegraphics[width=\columnwidth]{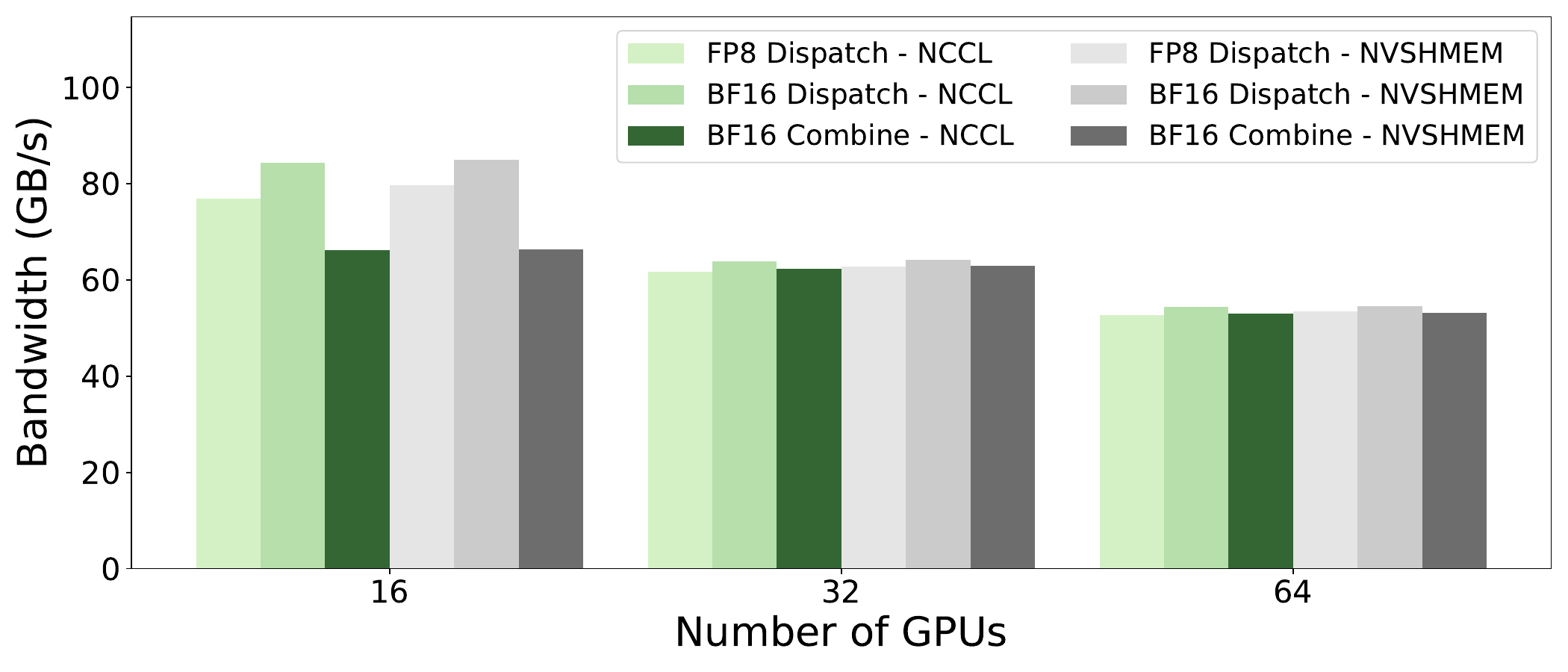}
    \caption{HT kernel bandwidth for NCCL GIN and NVSHMEM.}
    \label{fig:results:ht:bw}
\end{figure}

HT kernels optimize for MoE training and inference-prefill with 
large token batches (4096 tokens), using hierarchical communication where specialized SM roles 
(Senders, Forwarders, NVLink Receivers) minimize inter-node RDMA traffic while maximizing 
intra-node NVLink bandwidth. Figure~\ref{fig:results:ht:bw} presents dispatch and combine 
bandwidth for FP8 and BF16 precision across 2, 4, and 8 nodes, reporting separate RDMA and 
NVLink metrics.

Both implementations deliver comparable performance across all configurations. At 2 nodes 
(16 GPUs) with BF16 precision, dispatch operations achieve 84.36~GB/s RDMA bandwidth with NCCL GIN 
and 84.97~GB/s with NVSHMEM. At 8 nodes (64 GPUs), both implementations sustain 
approximately 53--54~GB/s RDMA bandwidth for dispatch operations. The results remain 
within 1--2\% across scales, precision modes, and operation types, demonstrating that 
NCCL GIN preserves HT throughput while enabling standardization on NCCL infrastructure.

\subsection{Low-Latency Kernels}
\label{sec:results:ll}

LL kernels (Section~\ref{sec:integration}) optimize for MoE inference-decode with small token 
batches (1--128 tokens), using full all-to-all RDMA mesh connectivity with per-expert signals 
and hybrid NVLink-RDMA paths. With BF16 precision and hidden dimension 7168, dispatch 
operations transfer 14,352-byte messages (14,336 bytes of token data plus 16 bytes of 
metadata for token source indexing), while combine operations transfer 14,336-byte messages. 
Zero-byte \texttt{put} with signal operations implement release-acquire semantics for data 
visibility. We evaluate under hybrid RDMA+NVLink and pure RDMA configurations to assess 
performance across deployment scenarios.

\begin{figure}[t]
    \centering
    \includegraphics[width=\columnwidth]{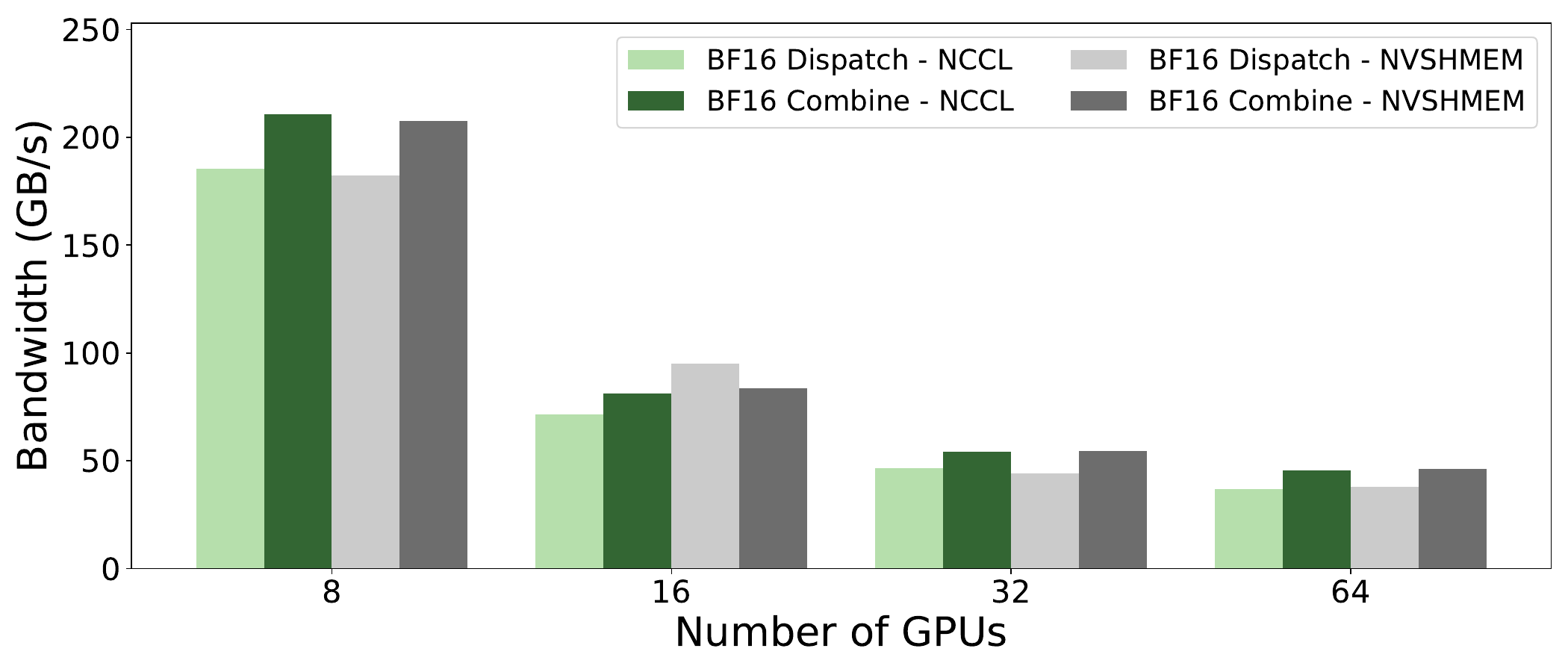}
    \caption{LL kernel bandwidth with NVLink enabled for NCCL GIN and NVSHMEM.}
    \label{fig:results:ll:nvlink:bw}
\end{figure}

\begin{figure}[t]
    \centering
    \includegraphics[width=\columnwidth]{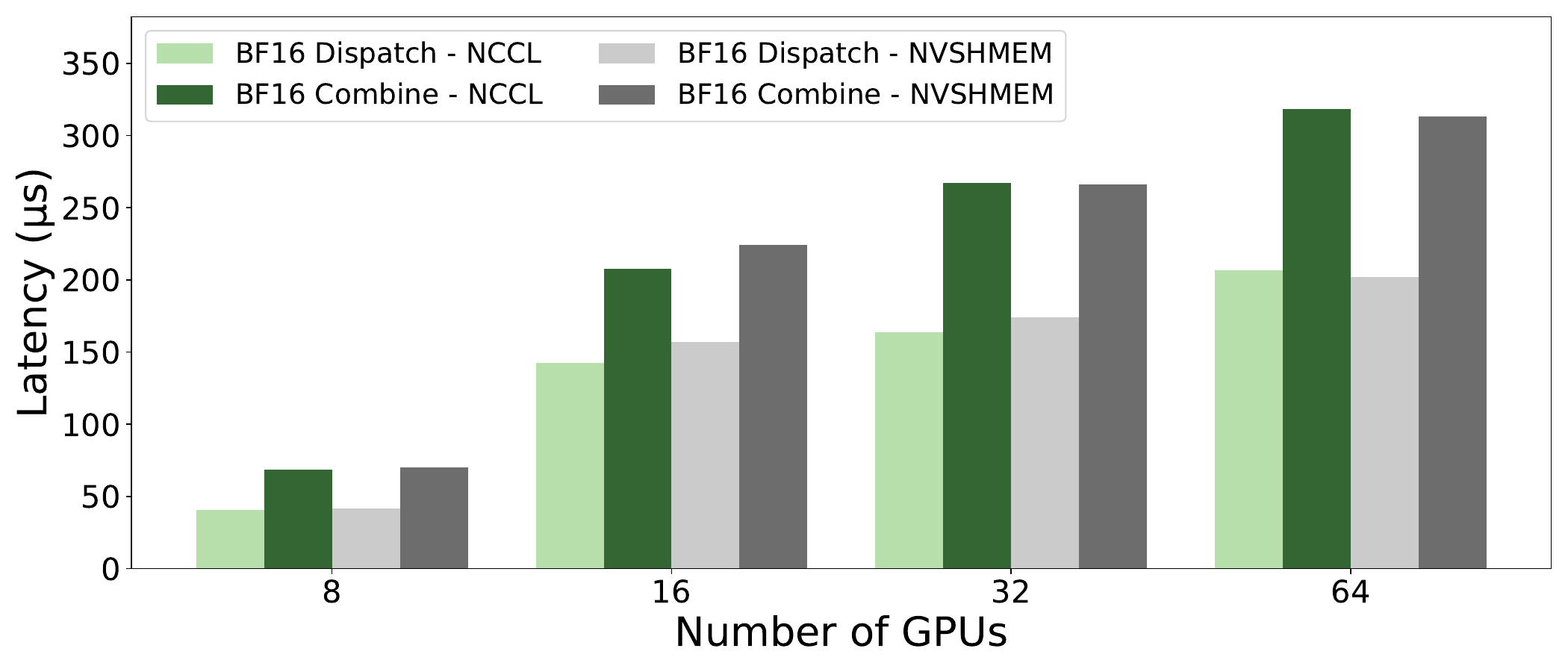}
    \caption{LL kernel latency with NVLink enabled for NCCL GIN and NVSHMEM.}
    \label{fig:results:ll:nvlink:lat}
\end{figure}

\noindent\textbf{LL Kernels with NVLink Enabled (RDMA+NVLink).}
This configuration represents typical deployments with intra-node NVLink and inter-node RDMA. 
Figures~\ref{fig:results:ll:nvlink:bw} and~\ref{fig:results:ll:nvlink:lat} show bandwidth 
and latency comparisons. At 1 node (8 GPUs), NCCL GIN slightly outperforms NVSHMEM: dispatch 
achieves 185.28~GB/s and 40.62~$\mu$s compared to 182.15~GB/s and 41.43~$\mu$s, while combine 
operations are virtually identical (211~GB/s, 69~$\mu$s).

At multi-node scales, both implementations show comparable performance with 
minor variations. NCCL GIN consistently delivers lower latency across scales 
(e.g., 9\% lower at 2 nodes: 142.51~$\mu$s compared to 157.00~$\mu$s). 
Combine operations remain within 1--3\% across all scales.

\begin{figure}[t]
    \centering
    \includegraphics[width=\columnwidth]{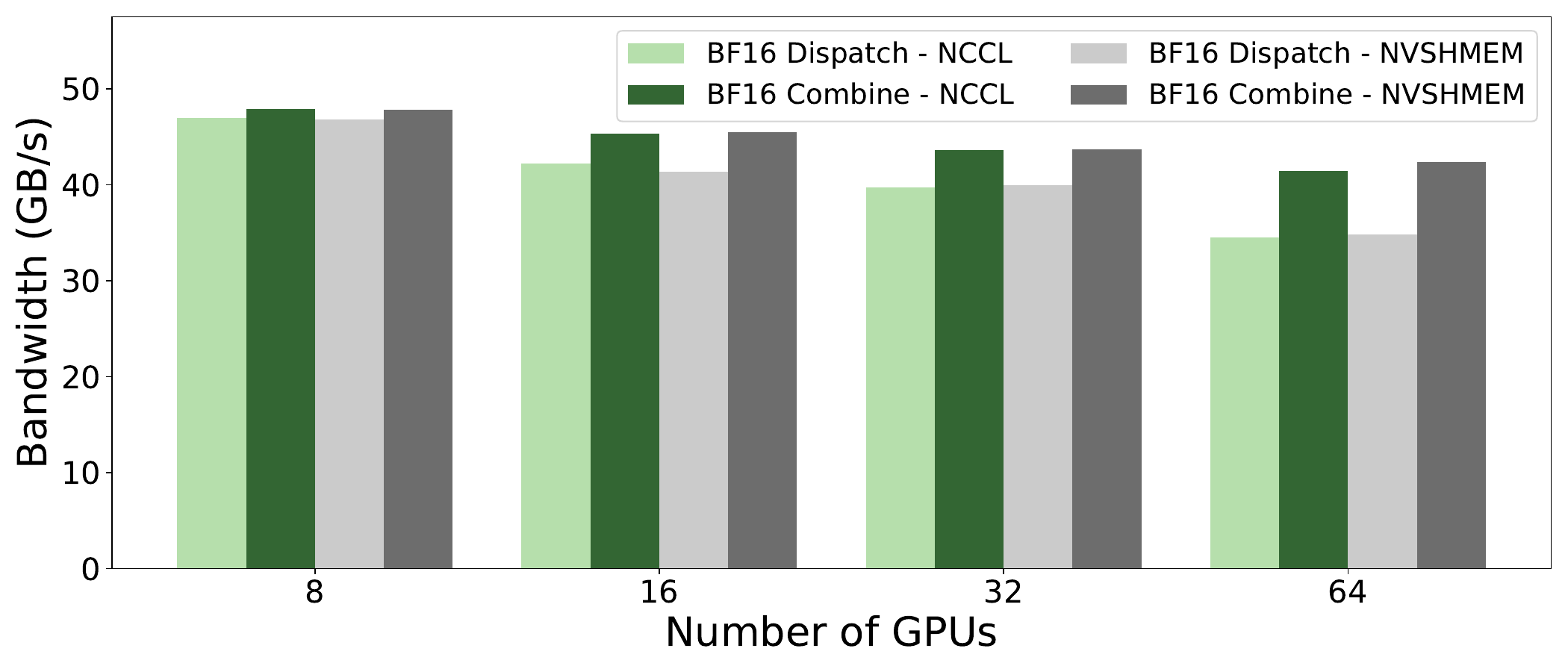}
    \caption{LL kernel bandwidth with NVLink disabled (pure RDMA) for NCCL GIN and NVSHMEM.}
    \label{fig:results:ll:rdma:bw}
\end{figure}

\begin{figure}[t]
    \centering
    \includegraphics[width=\columnwidth]{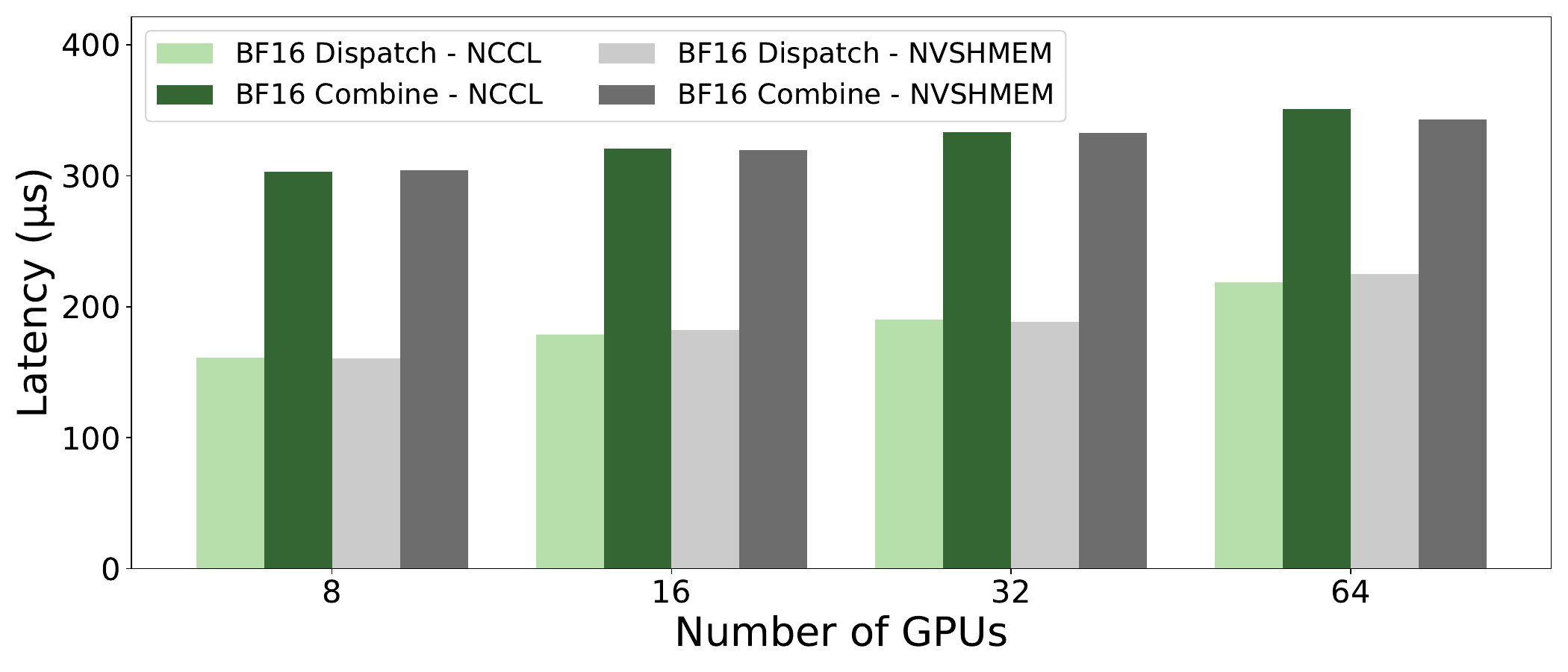}
    \caption{LL kernel latency with NVLink disabled (pure RDMA) for NCCL GIN and NVSHMEM.}
    \label{fig:results:ll:rdma:lat}
\end{figure}

\noindent\textbf{LL Kernels with NVLink Disabled (Pure RDMA).}
With NVLink disabled, all communication routes through RDMA—testing scenarios like cross-switch 
topologies or systems without NVLink. Figures~\ref{fig:results:ll:rdma:bw} 
and~\ref{fig:results:ll:rdma:lat} present measurements showing both implementations maintain 
comparable performance across scales, with most metrics within 1--2\%. At 1 node (8 GPUs), 
dispatch operations achieve 47.00~GB/s bandwidth and 160.82~$\mu$s latency with NCCL GIN, 
while NVSHMEM achieves 46.79~GB/s and 160.67~$\mu$s. At 8 nodes (64 GPUs), both 
implementations sustain approximately 34--35~GB/s bandwidth and 219--225~$\mu$s latency.

\subsection{Discussion}
The evaluation results demonstrate that NCCL GIN provides device-initiated 
communication capabilities with similar performance characteristics as NVSHMEM. Across 
microbenchmarks and application workloads (HT and LL kernels), GIN integrates 
device-initiated primitives with NCCL's topology-aware collectives within a 
single runtime, combining the flexibility of device-side APIs with NCCL's 
production infrastructure. The GIN implementation remains under 
active development, with planned optimizations including batching work queue entries and 
amortizing doorbell costs across multiple operations to further improve performance.

\section{Related Work}
\label{sec:related}

\noindent\textbf{Device-Initiated Communication Libraries.}
OpenSHMEM~\cite{chapman2010introducing,openshmem2012spec} established PGAS 
semantics for symmetric memory and one-sided operations, but early GPU extensions 
remained CPU-mediated~\cite{venkata2015exploring}. NVSHMEM~\cite{nvidia2023nvshmem,
potluri2017gpu} enabled device-callable operations from CUDA kernels, achieving 
60--75\% speedups by eliminating kernel launch overhead. However, NVSHMEM operates 
as a standalone runtime separate from existing collective communication frameworks. 
Early GPU-initiated RDMA efforts like GPUrdma~\cite{daoud2016gpurdma} and 
GIO~\cite{hamidouche2020gio} faced GPU-NIC memory consistency challenges, 
achieving 44\% improvements for irregular applications through kernel driver 
extensions. DOCA GPUNetIO~\cite{doca,doca-gpunetio} provides production-grade 
device-side RDMA APIs for InfiniBand and RoCE, forming the foundation for GIN's 
GDAKI backend.

\noindent\textbf{Collective Communication Runtimes.}
NCCL~\cite{nccl,Zhiyi25} provides topology-aware collective algorithms 
and production infrastructure for distributed training, traditionally using 
host-initiated communication. MPI RMA~\cite{mpi-forum} operations remain 
host-initiated, while UCX~\cite{ucx:15:hoti} and UCC~\cite{ucc} provide unified 
communication frameworks without device-callable primitives. NCCL 2.28 extends this 
with the Device API, integrating device-initiated capabilities (LSA, Multimem, GIN) 
into NCCL's existing infrastructure.

\noindent\textbf{MoE Communication Libraries.}
MoE architectures require irregular all-to-all routing with unpredictable message 
sizes~\cite{ren2024deepseekmoe}. DeepSpeed-MoE~\cite{rajbhandari2022deepspeed} 
employs hierarchical parallelism, while FasterMoE~\cite{he2022fastermoe} and 
Tutel~\cite{hwang2023tutel} optimize expert scheduling. DeepEP~\cite{deepep2025} 
and pplx-kernels~\cite{pplx-kernels} provide low-latency GPU-initiated primitives 
but operate independently from collective communication frameworks.

\noindent\textbf{GIN's Positioning.}
GIN uniquely integrates device-initiated network primitives into NCCL's production 
infrastructure through dual backends: GDAKI for direct GPU-to-NIC communication 
and Proxy for CPU-assisted operation on commodity hardware. This integration 
preserves NCCL's ecosystem compatibility while enabling device-driven communication 
for emerging workloads like MoE inference and kernel fusion patterns.

\section{Conclusions and Future Work}
\label{sec:conclusion}

Modern AI workloads---including MoE inference and compiler-generated fusion 
kernels---require direct GPU control over network operations, capabilities that 
extend beyond NCCL's traditional host-initiated model. This work introduces 
\textbf{GIN}, part of NCCL 2.28's Device API~\cite{nccl2.28}, which enables GPU threads to issue 
one-sided RDMA operations directly from CUDA kernels. GIN provides a unified 
three-layer architecture (host APIs, device APIs, and pluggable network backends with dual semantics) 
and supports both direct GPU-to-NIC communication via GDAKI and 
CPU-assisted operation on standard RDMA hardware.
Our evaluation validates GIN's practical viability: the GDAKI backend achieves 
16.7~$\mu$s round-trip latency for small messages, and DeepEP integration 
demonstrates competitive performance across DeepEP's High-Throughput and 
Low-Latency kernels with minimal code changes. 

\emph{Importantly, GIN's value 
extends beyond raw performance to ecosystem unification, delivering scalability 
and extensibility through integration with NCCL's production infrastructure}.
Applications gain 
access to a unified set of communication abstractions---Load/Store Accessible (LSA) 
for NVLink/PCIe, Multimem for NVLink SHARP, and GIN for network RDMA---enabling the 
right primitive for each pattern. Critically, this integration preserves NCCL's 
production-ready features: hierarchical communicators for multi-dimensional 
parallelism (expert-parallel, tensor-parallel, pipeline-parallel), fault tolerance 
and elasticity for resilient large-scale training, and topology-aware optimization. 
These capabilities eliminate the operational complexity of deploying multiple communication 
runtimes while providing the flexibility needed for emerging workloads like MoE 
inference and compiler-generated fusion kernels.

Future work will focus on the broader adoption of GIN in production applications such as 
PyTorch distributed training, TensorRT-LLM inference serving, vLLM and SGLang for LLM 
inference, and JAX/Triton compiler-generated kernels. We also plan to extend GIN's API 
with additional one-sided primitives to support emerging communication patterns and 
distributed algorithm requirements.

\section*{Acknowledgments}

The authors thank the NCCL and NVSHMEM teams for their contributions. The authors also thank Jeff Hammond and Matthew Nicely for their careful review of the manuscript and for their insightful feedback.

The authors acknowledge the use of Cursor AI to assist in the writing 
and editing of this manuscript. The authors reviewed and approved all 
content for accuracy and originality.

\bibliographystyle{IEEEtran}

\bibliography{references/deepepgin.bib}

\end{document}